\pdfoutput=1
\documentclass[twocolumn,10pt]{article}

\usepackage[letterpaper,margin=2.25cm]{geometry}
\usepackage[T1]{fontenc}
\usepackage{newtxtext,newtxmath}
\usepackage{amsmath}
\usepackage{graphicx}
\usepackage{booktabs}
\usepackage{makecell}
\usepackage{tabularx}
\usepackage{subcaption}
\usepackage{xcolor}
\usepackage{appendix}
\usepackage{float}
\usepackage{enumitem}
\usepackage{cite}
\usepackage{caption}
\usepackage{url}
\usepackage[colorlinks=false,pdfborder={0 0 0}]{hyperref}
\usepackage{balance}

\providecommand{\orcidlink}[1]{}
\providecommand{\authormail}[1]{\href{mailto:#1}{#1}}
\makeatletter
\@ifundefined{remark}{%
  \newcounter{remarkctr}%
    {\par\medskip}%
  \@addtoreset{remarkctr}{section}%
}{}
\makeatother
\providecommand{\listheading}[1]{\noindent\textbf{#1}\par\smallskip}

\usepackage{bm}
\usepackage{flushend}
\hypersetup{pdftitle={Sloshing Aware Tube MPC with Partial Tightening for Spacecraft Proximity Operations},
  pdfauthor={Omer Burak Iskender, Keck-Voon Ling, Jos\'{e} Luis Redondo Guti\'{e}rrez},
  pdfsubject={Preprint of IAC-26,C1,3,8,x108301}}

\usepackage{cuted}
\usepackage{dblfloatfix}
\begin{document}

\twocolumn[{%
\begin{center}
{\large\bfseries Sloshing Aware Tube MPC with Partial Tightening for Spacecraft Proximity Operations\par}
\vskip 0.93em
{\normalsize
\textbf{Omer Burak Iskender}\textsuperscript{a,b,*},\ \textbf{Keck-Voon Ling}\textsuperscript{b},\ \textbf{Jos\'{e} Luis Redondo Guti\'{e}rrez}\textsuperscript{c}\par}
\vskip 0.93em
{\normalsize\textsuperscript{a}\,\textit{Final Proximity Space Systems}\par}
\vskip 0.31em
{\normalsize\textsuperscript{b}\,\textit{School of Electrical and Electronic Engineering, Nanyang Technological University, Singapore. {E-mail:~\authormail{iske0001@e.ntu.edu.sg},~\authormail{ekvling@ntu.edu.sg}}}\par}
\vskip 0.31em
{\normalsize\textsuperscript{c}\,\textit{Institute of Space Systems, German Aerospace Center (DLR), Bremen, Germany. {E-mail:~\authormail{jose.redondo-gutierrez@dlr.de}}}\par}
\vskip 0.31em
{\normalsize\textsuperscript{*}\,Corresponding author\par}
\vskip 0.62em
\noindent\rule{\linewidth}{0.4pt}
\vskip 0.25em
{\small\itshape \textbf{Presented at} the 77th International Astronautical Congress (IAC 2026), Antalya, T\"{u}rkiye, 5--9 October 2026; paper IAC-26,C1,3,8,x108301. This is the authors' preprint, revised after the congress to carry the results as presented there; it is not the version of record.\par}
\vskip 0.25em
\noindent\rule{\linewidth}{0.4pt}
\end{center}
\vskip 0.62em
\begin{center}
\textbf{Abstract}
\end{center}
\noindent
This paper presents a sloshing-aware tube model predictive control (MPC) for spacecraft proximity operations, which sizes its constraint tightening from the propellant's slosh reaction and applies it only to the constraints that bind, the approach corridor and a pointing budget about a fixed inertial attitude. A chaser closing on a Mars Sample Return interface carries propellant that sloshes relative to its hub, and a tube MPC certified against rigid-body disturbances cannot tell whether that motion matters. Earlier work flew the reference guidance in the General Mission Analysis Tool, and we fly the same elliptical-orbit rendezvous in Basilisk, a six-degree-of-freedom multibody simulator that represents the propellant with coupled slosh particles. Once a burn-delivery defect in the simulation loop is corrected, the rigid Basilisk plant reproduces the earlier campaign. Flown unchanged from homing to contact on the sloshing plant, the reference guidance's hub-point miss grows on every seed, at almost no extra propellant. A frozen-liquid control suggests that most of this loss comes from the uncontrolled hub tumbling about the propellant's offset centre of mass rather than from the liquid itself, and a tube ablation on the same guidance is inconclusive at this sample size. Several controllers are then compared on the terminal and closing legs to isolate the slosh disturbance. Sizing the attitude tube's tightening from the slosh torque lowers the settled error of a fixed-attitude hold by $12.8\%$ over both nominal MPC and tube MPC, in a replay without orbital motion, while the same tightening in translation changes nothing measurable. A horizon-selecting variant saves propellant but arrives more slowly and leaves the corridor more often during the blind coast. On the docking geometry, an extended slosh-aware MPC lets the liquid past its linear limit on 8 of 2400 approaches against 674 for nominal MPC, points better in all 24 campaigns, and docks 300 of 300 approaches against 289 with 0.5~N\,m reaction wheels.
\vskip 1.55ex
}]
\begin{strip}
\noindent
\begin{minipage}[t]{0.47\textwidth}
\listheading{Nomenclature}
\noindent\small
$m_s$ \quad participating (slosh) mass, one particle \\
$\rho,\dot\rho,\omega_s$ \quad slosh displacement, rate, natural frequency \\
$L,\ell_{\mathrm{tank}}$ \quad slosh torque lever arm, tank offset \\
$\bar e_j,\bar w$ \quad tube half-width at step $j$, disturbance bound \\
$K,A_{\mathrm{cl}}$ \quad ancillary gain, closed loop $A+BK$ \\
$N,N^\star,T_s$ \quad horizon, selected horizon, sampling period \\
$\theta_{\max},\alpha$ \quad pointing budget, corridor half-angle
\normalsize
\end{minipage}\hfill
\begin{minipage}[t]{0.47\textwidth}
\listheading{Acronyms/Abbreviations}
\noindent\small
CW, YA: Clohessy--Wiltshire, Yamanaka--Ankersen \\
HD-TUBE: horizon-selecting tube MPC (Sec.~\ref{sec:ctrl}) \\
HDE, HDE-NT: element-space OSTG, tightening on / off \\
INTG: impulsive nominal translational guidance (closing) \\
MPC, MSR: model predictive control, Mars Sample Return \\
OSTG: orbit synchronisation guidance (homing) \\
RTN, TOF: radial-transverse-normal, target-orbit frame
\normalsize
\end{minipage}
\end{strip}

\section{Introduction}
\label{sec:intro}

Propellant slosh affects a spacecraft in four ways. The liquid puts forces and torques on the vehicle, the thrusters spend propellant cancelling the oscillation, the free surface can uncover a tank outlet, and the liquid loads flexible structures near resonance \cite{Abramson1966,Dodge2000,Ibrahim2005,MacesHernandez2023}. This study measures the first two, as the error of a fixed-attitude hold, the hub-point miss at contact and the propellant spent on them; the feed line and flexible appendages return as future work in Section~\ref{sec:conclusions}.

A tube-based model predictive controller (MPC) for rendezvous is usually certified against a rigid hub, with its tightening sized from unmodelled acceleration, actuator error or mass uncertainty. The liquid in a partly full tank on a Mars Sample Return (MSR) chaser moves relative to the hub and pushes back with a force and a torque that the rigid model never predicts. Only a coupled plant can show whether the rigid certificate still protects the vehicle.

This paper continues one line of work. Its first steps were MPC with constraint tightening for docking with uncooperative targets \cite{iskender2018constraints,Iskender2019,burak2020model} and a horizon-dependent tightening on elliptical orbits with a conditional robust constraint-satisfaction guarantee \cite{Iskender2026EllipticalTube,iskender2026horizon}. The companion study \cite{IskenderActa2026} appended the slosh states to the predictor, but at the $3$~s sampling period of a docking approach a box tube on those states diverges (Section~\ref{sec:ctrl}). The controller here therefore keeps the rigid predictor, sizes its tightening from a bound on the reaction force and torque the propellant can exert, and applies it only to the two safety constraints of a docking approach, the approach corridor and a pointing budget, here about a fixed inertial attitude. That restriction is the partial tightening of the title (Eqs.~\eqref{eq:ocp}--\eqref{eq:attocp}).

The scenario is Hartley's elliptical MSR rendezvous ($e=0.204$, $a=4643$~km) \cite{Hartley2012}. It was flown earlier in the General Mission Analysis Tool (GMAT) \cite{Hughes2014GMAT}, and here on a six-degree-of-freedom Basilisk plant \cite{Kenneally2020} whose lateral slosh modes are particles coupled to the hub, after the rigid plant was validated against the GMAT run. The first comparison flies the reference guidance \cite{Iskender2026EllipticalTube,iskender2026horizon} unmodified from the homing hand-over to contact, rigid against coupled plant and tightening on against off, twenty paired seeds each. The second compares proportional-derivative (PD) control, a linear-quadratic regulator (LQR), nominal MPC, a tube sized for a rigid-body allowance, the slosh-aware tube and a horizon-selecting tube (HD-TUBE) on $100$ paired trials from $98$~m to contact, a $100$-seed attitude campaign that replays the final-approach burns without orbital motion and $20$ paired trials over the $4393$~m closing leg. A further $24$ campaigns fly the docking geometry with a $400$~kg tank and two vehicles from the slosh literature.

Two results stand out. On the full mission the reference guidance's median hub-point miss is $0.1067$~m on the coupled plant against $0.0005$~m on the rigid one, worse on every seed, but a frozen-liquid control traces most of that loss to the uncontrolled hub tumbling about the propellant's offset centre of mass. On the terminal leg, sizing the attitude tube from the slosh torque lowers the median settled error of a fixed-attitude hold by $12.8\%$ and the ninety-fifth percentile by $14.1\%$ against nominal and rigid-tube MPC, on $99$ of $100$ paired seeds, in a replay without orbital motion (Table~\ref{tab:attitude}).

\subsection*{Related work}

Tube MPC \cite{Langson2004,Mayne2005,Rakovic2005} is established for spacecraft proximity operations \cite{Weiss2015,Hartley2015,Hartley2012,Mammarella2018,Buckner2018,Eren2017}, with tube widths sized from actuator error or unmodelled acceleration. Work on pursuit--evasion games \cite{iskender2026convex} and on safe-start regions for a tumbling target \cite{Iskender2026IAC} likewise places the uncertainty outside the chaser's own propellant. Slosh dynamics are classical \cite{Abramson1966,Dodge2000,Satterlee1964}, and slosh-aware control has been studied for launch and landing vehicles \cite{Fari2022,Vreeburg2005,Ibrahim2005}, but rarely tied to a docking-phase tube. The companion study \cite{IskenderActa2026} analyses the coupled linear model. Here we fly the multibody plant that model approximates and take the disturbance bound from the plant instead of assuming it.

\section{Relative Motion and Coupled Slosh Dynamics}
\label{sec:dyn}

\subsection*{Predictors and frames}

The target follows Hartley's MSR ellipse \cite{Hartley2012}, so the mission-scale predictor is the Yamanaka--Ankersen (YA) state-transition matrix \cite{Yamanaka2002,Tschauner1965}; at this eccentricity the constant-coefficient Clohessy--Wiltshire (CW) model diverges from a nonlinear truth by a factor of $280$ over one orbit \cite{Iskender2026EllipticalTube,iskender2026horizon}. For the six-controller legs we chose the predictor against the tube's one-step half-width of $1.38$~mm. At the terminal leg's dispersed initial states and $T_s=3$~s, CW's one-step position error against a Kepler propagation of both vehicles is $0.805$~mm at the median and $1.03$~mm at worst, and YA's is five orders smaller, so the terminal controllers use CW. At the $4393$~m closing entry CW errs by $3.43$~mm and YA by $0.030$~mm, so the closing controllers use YA.

The controllers work in the rotating target-orbit frame (TOF), whose rows are $[\hat{\mathbf t},-\hat{\mathbf h},-\hat{\mathbf r}]$ in the radial--transverse--normal (RTN) triad, so that $\mathbf x_{\mathrm{YA}}=-\mathbf x_{\mathrm{TOF}}$ component for component. Every state handed to a predictor carries the rotating-frame velocity $\dot{\mathbf d}_{\mathrm{inertial}}-\bm\omega\times\mathbf d$. Omitting it would move the predicted state by $0.30$~m in one step at the $98$~m hand-over, more than the whole capture envelope.

\subsection*{The coupled plant}

\begin{figure}[tbp]
\centering
\includegraphics[width=\columnwidth]{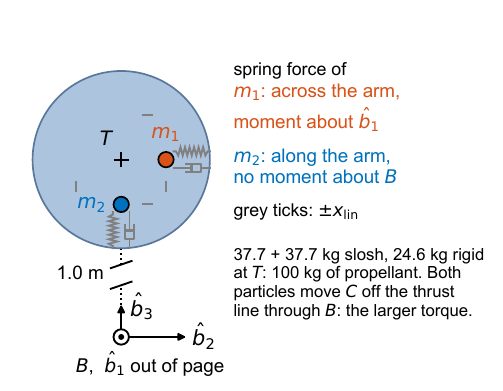}
\caption{The tank seen down $\hat{\mathbf b}_1$: tank centre $T$, body origin $B$, particles $m_1$ and $m_2$ on slosh axes $\hat{\mathbf b}_2$ and $\hat{\mathbf b}_3$, displacements drawn at $3.5$ times the linear limit.}
\label{fig:endview}
\end{figure}

\begin{figure*}[tbp]
\centering
\includegraphics[width=\textwidth]{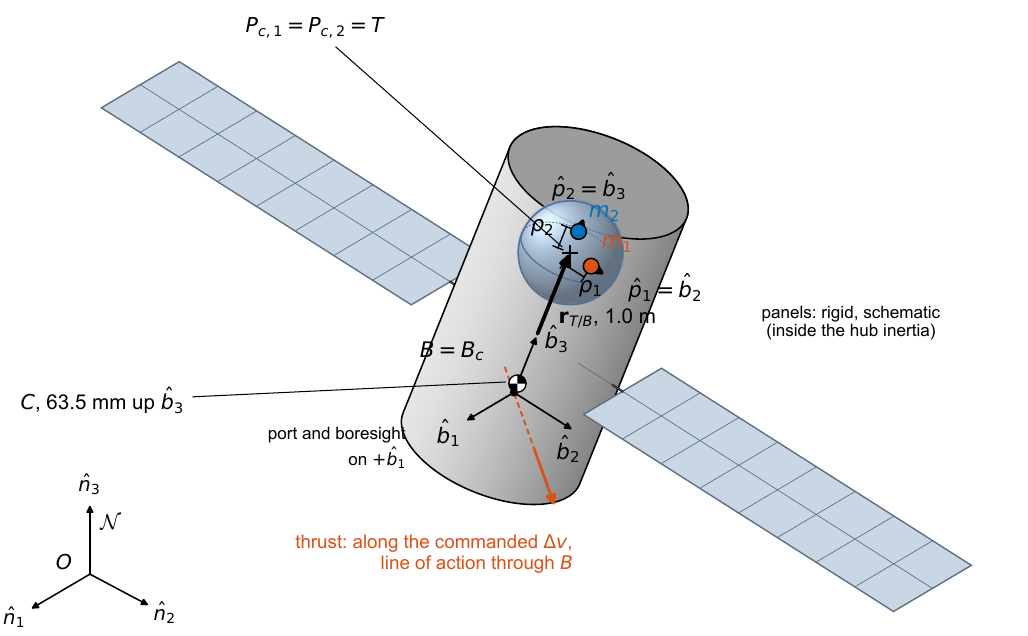}
\caption{The coupled vehicle: body origin $B$, tank centre $T$, particles $m_1$ and $m_2$, system centre of mass $C$ and the thrust line through $B$. Hub and panels are schematic.}
\label{fig:vehicle}
\end{figure*}

A rigid-body plant cannot represent propellant that moves relative to the hub, so every campaign is flown on a Basilisk six-degree-of-freedom plant. Its tank is represented by translating slosh particles, integrated as state effectors and coupled to the hub by Basilisk's Newton--Euler back-substitution \cite{Kenneally2020}. Table~\ref{tab:vehicle} lists the vehicle, tank and scenario. The nominal $100$~kg fill, $6.3\%$ of the vehicle mass, is an assumption of this study. Its first antisymmetric lateral mode participates with $m_s=37.7$~kg per lateral direction and is carried as two particles on the slosh axes $\hat{\mathbf b}_2$ and $\hat{\mathbf b}_3$, both at rest at the tank centre $T$, $1$~m up $\hat{\mathbf b}_3$ from the body origin $B$ through which the thrust acts (Figs.~\ref{fig:endview} and~\ref{fig:vehicle}). The remaining $24.6$~kg sits at $T$, and the hub has Hartley's mass and inertia \cite{Hartley2012} and no hinged panels.

Two plant properties make the comparison with the reference campaign possible. The mission campaigns fly GMAT's own $50\times50$ Mars field with the field's $\mu$, as GMAT does ($0.015\%$ from the guidance's), and the six-controller legs fly a point-mass field. Basilisk's dynamics task also advances only at multiples of its period, so a burn shorter than one period, such as a $0.445$~s final-approach command on the $0.5$~s step, was dropped outright. We now spread every burn over a whole number of integrator steps at a force scaled to deliver the exact impulse, and across the $1440$ trials of the six-controller campaigns the ratio of delivered to deliverable $\Delta v$ is $1.000000$ in every trial.

\begin{table}[tbp]
\centering
\caption{Vehicle, tank and scenario parameters as flown on the Basilisk plant, with the reference guidance's per-phase sampling and horizon.}
\label{tab:vehicle}
\footnotesize
\setlength{\tabcolsep}{3pt}
\begin{tabularx}{\columnwidth}{@{}>{\raggedright\arraybackslash}p{0.25\columnwidth}>{\raggedright\arraybackslash}X@{}}
\toprule
Quantity & As flown \\
\midrule
\multicolumn{2}{@{}l}{\emph{Vehicle}} \\
Mass, inertia & 1575~kg, $\mathrm{diag}(1450,1800,1200)$~kg\,m$^2$ \cite{Hartley2012}; the rigid plant holds all of it in the hub at $B$, with no tank \\
Thrusters & $8\times22$~N on the body diagonals (layout assumed); 36.09~N net along the commanded $\Delta v$, line of action through $B$, impulse delivered exactly \\
Charged $\Delta v$, propellant & $1.805\times$ the commanded $|\Delta v|$ (six-controller legs) or $1.805\times$ its eight-thruster allocation (mission); rocket equation from 1575~kg at $I_{sp}=220$~s (assumed) \\
Attitude actuation & body torque, no thruster moment; ball of 20~N\,m, 4~N\,m in the attitude campaign; held attitude fixed in inertial space; no torque applied in the mission runs \\
\midrule
\multicolumn{2}{@{}l}{\emph{Tank and slosh}} \\
Tank & sphere of radius 0.30~m, centre $T$ 1~m along $\hat{\mathbf b}_3$ from $B$; modal analogue: cylinder of the same radius, liquid depth 0.352~m at nominal fill \\
Propellant & 100~kg nominal (assumed), sphere capacity 113.6~kg; fill multiplier clipped to $[0.75,1.45]$; the mission runs fly 1.00 \\
Particles & 2, slosh axes $\hat{\mathbf b}_2$ and $\hat{\mathbf b}_3$, equilibrium at $T$; $m_s=37.69$~kg each and 24.61~kg rigid at nominal fill (35.80 to 38.56~kg and 3.41 to 67.88~kg over the clip) \\
First mode & burn-regime spring on every leg: period 16.36~s, damping ratio 0.0053, $x_{\mathrm{lin}}=43.3$~mm \\
\midrule
\multicolumn{2}{@{}l}{\emph{Scenario}} \\
Target orbit & $a=4643$~km, $e=0.20441$, $i=15^\circ$ \\
Gravity & mission: GMAT's $50\times50$ Mars field; six-controller legs: point-mass Mars; attitude campaign: none \\
Reference guidance & from 476.8~km: homing at 600~s, $N\in[8,50]$ chosen each step; closing at 300~s, $N=20$; final approach at 3~s, $N=15$ \\
Six-controller legs & terminal from 98~m (centre to centre) to contact; closing from 4393~m to a 100~m gate; $T_s=3$~s, $N=25$; HD-TUBE $N\in\{15,25,35,45,60,75\}$ \\
Interface & hub points 5.65~m apart along-track at contact (port arms 2.5 and 3.15~m); blind below 3~m; capture within 0.20~m of the along-track axis (assumed); corridor $3^\circ$ ($1.5^\circ$ to $10^\circ$ in the sweep) \\
\bottomrule
\end{tabularx}
\end{table}

\subsection*{Slosh model and disturbance bounds}

Equivalent models of slosh range from the linear mass--spring--damper analogue \cite{Fries2012,IskenderActa2026} through nonlinear and multibody pendulums \cite{Ibrahim2005,MacesHernandez2023} to the effector mass-matrix formulation flown here \cite{Kenneally2020}. That formulation solves the particles and the hub together, so the reaction on the hub is exact and momentum is conserved by construction. The particle parameters and the disturbance bounds below come from the mass--spring--damper analogue, which the companion study carries inside its predictor.

For a cylindrical tank the first antisymmetric mode's frequency and participating mass follow the classical analogue \cite{Abramson1966,Dodge2000}, with $\xi_1=1.8412$ the first root of $J_1'(\xi)=0$. The plant keeps the spring $k=m_s\omega_s^2$ at the burn's specific force on every leg, which puts the first-mode period at $16.4$~s, $0.18$ cycles per $3$~s terminal step against $36.7$ per $600$~s homing step and $18.3$ per $300$~s closing step. Only on the terminal leg does the controller see the mode rather than an aliased disturbance, so that leg carries the controller comparison.

The particle's reaction on the hub about $B$ has a term in $\ddot\rho$ and a Coriolis term in $\dot\rho$,
\begin{equation}
  \mathbf F_{\mathrm{slosh}} = -\big(m_s\hat{\mathbf p}\,\ddot\rho + 2[\tilde{\bm\omega}]m_s\dot\rho\,\hat{\mathbf p}\big),
  \label{eq:fslosh}
\end{equation}
but most of the torque about $B$ comes from the thrust acting on the centre of mass that the displaced particle has moved off the thrust line,
\begin{equation}
  \tau_{\mathrm{com}} = F_{\mathrm{thrust}}\,\frac{m_s}{m_{\mathrm{tot}}}\,L,
  \qquad
  \tau_{\mathrm{reac}} = m_s\,L\,\rho_{\max}\,\omega_s^2,
  \label{eq:tslosh}
\end{equation}
with $L=\ell_{\mathrm{tank}}+\rho_{\max}$ the tank's offset from $B$ plus the particle's displacement. A tank at the body origin would give a bound of only $0.048$~N\,m, so the tank flown here sits $\ell_{\mathrm{tank}}=1$~m out, an assumption of this study. There the two terms bound to $0.90$ and $0.25$~N\,m, $1.15$~N\,m together, and the translational reaction is $0.24$~N on the $1575$~kg vehicle, or $1.5\times10^{-4}$~m/s$^2$. The same arm carries a static term that the bound leaves out. At rest the propellant sits at $T$, which puts the system centre of mass $63.5$~mm up $\hat{\mathbf b}_3$ from $B$, so a lateral burn at the $36.09$~N net thrust carries $2.29$~N\,m about it whether or not the liquid moves. That is $1.4$~N\,m more than the bound's one-particle share, and the rigid plant, with all $1575$~kg at $B$, has no such arm. The six-controller campaigns hold the hub at a fixed inertial attitude against this torque, so body axis $\hat{\mathbf b}_1$, which carries the port, is $54.8^\circ$ from the along-track approach line at the $98$~m hand-over and $67$ to $73^\circ$ from it at contact; the attitude campaign holds a fixed frame without orbital motion, and the mission campaigns command none and the hub spins (Section~\ref{sec:homing}).

We checked the plant twice before trusting any result. With the dashpot disabled and no external force, the rotational angular momentum about the system centre of mass and the rotational energy drift by $2.6\times10^{-15}$ and $2.8\times10^{-12}$ in relative terms over $60$~s. A case repeated in a field-free environment and in Mars orbit changes the particle displacement by $8.4\times10^{-17}$~m, so the restoring force in coast is the particle's spring alone. The analogue has limits of its own. As the tank empties, the participating mass fraction tends to $2/(\xi_1^2-1)=0.837$, and with one particle per lateral direction a diagonal excitation counts the mode twice, so the analogue is conservative by up to a factor of two. Below a fill multiplier of $0.70$, $2m_s$ exceeds the propellant mass, so the dispersed campaigns clip the multiplier to $[0.75,1.45]$ and the mission flies the nominal fill. Above $1.14$, or $113.6$~kg of hydrazine, the sphere of $0.30$~m radius would overflow, and $23$ of $100$ terminal, $28$ of $100$ attitude and $4$ of $20$ closing seeds exceed it without either model objecting. The tank does not deplete either, whereas a $100$~kg tank fed from homing onwards would be at $0.50$ by the closing hand-over, below the $0.70$ edge.

\section{Controllers}
\label{sec:ctrl}

\subsection*{The reference guidance}

The mission comparison flies the reference architecture's guidance \cite{Iskender2026EllipticalTube,iskender2026horizon} unmodified. Homing runs an element-space orbit-synchronisation program (OSTG), the closing leg the Cartesian tube of this section (INTG) and the terminal approach the same tube coupled to attitude, with the tightening recursion of Eq.~\eqref{eq:tube} carried through all three phases. The published form flies OSTG with its tightening on, closes with a reduced-tracking coast that restores full tracking inside $250$~m so that the vehicle arrives with its closing rate nulled, and solves the terminal translation and attitude in one program, although the mission runs apply no torque (Table~\ref{tab:vehicle}). The tube ablation compares HDE, the element-space OSTG with its tightening on, against HDE-NT, the same program with it off.

\subsection*{Six controllers on the closing and terminal legs}

The six controllers share one cost and one corridor, so the comparison isolates the slosh disturbance. On the terminal leg all six act in the TOF frame on the CW pair and close the loop every $T_s=3$~s down to a range of $3$~m. Below that range no controller acts and the vehicle coasts blind to contact. The closing leg reuses this structure from $4393$~m to a $100$~m gate on the YA pair, tracking a glide-slope velocity reference ($\bar v=2$~m/s and $a_z=0$ in Eq.~\eqref{eq:ocp}) with no cost on along-track position. PD and LQR use the weights of Eq.~\eqref{eq:ocp} and are clipped to the same per-axis $\Delta v$ box. The corridor is imposed only on these two legs, whose hub-point state, measured from a $5.65$~m along-track standoff, is what the constraint assumes; imposed on homing, where the state is a tracking error about a drift arc, it made the guidance infeasible on five of seven steps.

\subsection*{How the slosh enters the controller}

A tube can be made slosh-aware in two ways. The companion study \cite{IskenderActa2026} appends the two lateral particles to the predictor, ten states coupled through the centre-of-mass identity $\mathbf x_{\mathrm{com}}=\mathbf x+E\mathbf x_s$, with $E$ carrying the mass share $\mu_s=m_s/m_{\mathrm{tot}}$ on the lateral rows, so that momentum is conserved by construction,
\begin{equation}
\begin{bmatrix}\mathbf x\\ \mathbf x_s\end{bmatrix}_{k+1}
=\begin{bmatrix}A & AE-EA_s^d\\ 0 & A_s^d\end{bmatrix}
\begin{bmatrix}\mathbf x\\ \mathbf x_s\end{bmatrix}_{k}
+\begin{bmatrix}B-EB_s\\ B_s\end{bmatrix}\mathbf u_k,
\label{eq:aug}
\end{equation}
with $A_s^d$ the exact discretisation of the particles and $B_s$ their response to a burn. A box tube on that model must carry the slosh partition through $|A_s^d|$. The mode is stable, $\rho(A_s^d)=0.994$, yet $\rho(|A_s^d|)=1.31$ at $\omega_dT_s=1.15$~rad and still $1.25$ at the analogue's coast frequency. A box cannot represent phase, so each rotation, although an isometry, inflates it by up to $\sqrt2$, and this wrapping effect of interval arithmetic \cite{Moore1966} makes the tube diverge although nothing is unstable. Tightening only the rigid partition, as \cite{IskenderActa2026} does, needs a separate bound on the slosh error, which a box recursion cannot supply either.

The controllers flown here keep the rigid predictor, six states in translation and nine in attitude, and bound the slosh partition in advance instead. The analogue holds only while the particle stays within its linear limit, $|\rho|\le x_{\mathrm{lin}}$ with $|\dot\rho|\le\omega_sx_{\mathrm{lin}}$, so the reaction it can put on the hub is bounded by Eqs.~\eqref{eq:fslosh} and~\eqref{eq:tslosh} evaluated there. That bound enters the rigid recursion as a constant disturbance, $a_w$ in translation and $\bar w_\tau$ in attitude, and the tube stays finite. The tightening is also partial in the constraints it touches, which are only the corridor rows and the pointing budget. Tightening the predicted inputs by $|K|\bar{\mathbf e}_j$ as well would remove at most $5.4\%$ of $u_{\max}$ in translation, and the applied input, the program's first move, satisfies its bound directly. With the slosh torque bound the tightened budget $\bar\theta_j$ on the attitude-hold error of Eq.~\eqref{eq:attocp} falls to its $0.05^\circ$ floor from the fourth prediction step, below the hold error nominal MPC settles to, whereas the rigid allowance leaves it above $0.81^\circ$ over the whole horizon.

\subsection*{The slosh-aware tube MPC problem}

Every MPC variant of the comparison solves, at each step $k$, one quadratic program on the deviation $\delta\mathbf x=\mathbf x-\mathbf x^{\mathrm{ref}}$ of the hub-point relative state $\mathbf x=[x,y,z,\dot x,\dot y,\dot z]^\top$ from a $5.65$~m along-track standoff (along-track, cross-track, radial, in the TOF frame) from a reference $\mathbf x^{\mathrm{ref}}=[0,0,a_z,-v^{\mathrm{ref}},0,0]^\top$ that holds the radial aim-off $a_z$ and the approach speed $v^{\mathrm{ref}}=\min(\bar v,\,k_v x)$, both frozen over the horizon:
\begin{align}
\min_{\mathbf u_0,\dots,\mathbf u_{N-1}}\ \ & \sum_{j=1}^{N}\|\delta\mathbf x_j\|_Q^2+\sum_{j=0}^{N-1}\|\mathbf u_j\|_R^2 \label{eq:ocp}\\
\text{s.t.}\ \ & \delta\mathbf x_{j+1}=A\,\delta\mathbf x_j+B\,\mathbf u_j,\quad \delta\mathbf x_0=\mathbf x_k-\mathbf x^{\mathrm{ref}},\nonumber\\
& |\delta y_j|\le\tan\alpha\;\delta x_j-\bar e^{\,y}_j,\nonumber\\
& |\delta z_j|\le\tan\alpha\;\delta x_j-\bar e^{\,z}_j, \label{eq:cone}\\
& |u_{j,i}|\le u_{\max},\quad i=1,2,3,\nonumber
\end{align}
with the corridor rows applied at $j=1,\dots,N$ and the input bound at $j=0,\dots,N-1$. The terminal state $\delta\mathbf x_N$ carries the stage weight $Q$, with no separate terminal cost. Here $A$ is the exact transition matrix of CW (terminal leg) or YA (closing leg) over $T_s=3$~s, and $B=A\,[\,\mathbf 0\ \ I_3\,]^\top$ applies the $\Delta v$ impulse at the start of each step; $N=25$, $Q=\mathrm{diag}(0.3,100,100,800,800,800)$, $R=6000\,I_3$, $u_{\max}=F_{\mathrm{net}}T_s/m=0.0687$~m/s, $\alpha=3^\circ$ and $(\bar v,k_v)=(0.05~\mathrm{m/s},\,0.004~\mathrm{s}^{-1})$ on the terminal leg. Lateral position is weighted far above along-track because the hub point's lateral offset at the standoff is the scored miss, whereas an along-track error only makes the arrival early or late. The corridor is the square that circumscribes the cone, an outer approximation admitting up to $\sqrt2\tan\alpha\,\delta x_j$ on a diagonal, and violations are counted against the true cone. The half-widths follow the box recursion
\begin{equation}
\bar{\mathbf e}_{j+1}=|A_{\mathrm{cl}}|\,\bar{\mathbf e}_j+|B\mathbf 1_3|\,a_wT_s,\qquad \bar{\mathbf e}_0=\mathbf 0,
\label{eq:tube}
\end{equation}
with $A_{\mathrm{cl}}=A+BK$ and $K$ the discrete algebraic Riccati gain for $(A,B,Q,R)$. The variants differ only in the disturbance-acceleration bound $a_w$, which is $0$ for nominal MPC, $10^{-6}$~m/s$^2$ for the rigid tube and, for the slosh-aware tube, the reaction bound of Eq.~\eqref{eq:fslosh} at the linear limit,
\begin{equation}
a_w=\frac{m_s\big(\omega_s^2\rho_{\max}+2\,\omega_b\,\dot\rho_{\max}\big)}{m_{\mathrm{tot}}}=1.5\times10^{-4}~\mathrm{m/s}^2,
\label{eq:aw}
\end{equation}
with $\rho_{\max}=x_{\mathrm{lin}}$, $\dot\rho_{\max}=\omega_s\rho_{\max}$ and $\omega_b=2\times10^{-4}$~rad/s the hub rate the attitude loop holds. An infeasible program falls back to the ancillary law $\mathbf u=K\,\delta\mathbf x$, clipped to the input box, and the fallback is counted.

The attitude channel poses the same problem on the attitude error about a fixed inertial reference, with integral action, $\boldsymbol\xi=[\boldsymbol\theta^\top,\boldsymbol\omega^\top,\boldsymbol\eta^\top]^\top$ with $\dot{\boldsymbol\eta}=\boldsymbol\theta$:
\begin{align}
\min_{\boldsymbol\tau_0,\dots,\boldsymbol\tau_{N_a-1}}\ \ & \sum_{j=1}^{N_a}\|\boldsymbol\xi_j\|_{Q_a}^2+\sum_{j=0}^{N_a-1}\|\boldsymbol\tau_j\|_{R_a}^2 \label{eq:attocp}\\
\text{s.t.}\ \ & \boldsymbol\xi_{j+1}=A_a\boldsymbol\xi_j+B_a\boldsymbol\tau_j,\quad \boldsymbol\xi_0=\boldsymbol\xi_k,\nonumber\\
& |\theta_{j,i}|\le\bar\theta_j,\qquad \boldsymbol\tau_j\in\mathcal T,\nonumber
\end{align}
with the pointing rows applied at $j=1,\dots,N_a$ and $\boldsymbol\tau_j\in\mathcal T$ at $j=0,\dots,N_a-1$. Here $(A_a,B_a)$ is the zero-order hold of $\dot{\boldsymbol\theta}=\boldsymbol\omega$, $J\dot{\boldsymbol\omega}=\boldsymbol\tau$ at $T_s=3$~s, $Q_a=\mathrm{diag}(4\times10^4I_3,\,2\times10^3I_3,\,800\,I_3)$, $R_a=I_3$, $N_a=12$, and $\mathcal T$ is a $26$-direction polytope inscribed in the ball $\|\boldsymbol\tau\|\le\tau_{\max}$. The tightening acts on the pointing budget, $\bar\theta_j=\max(\theta_{\max}-\max_i\bar e_{j,i},\,0.05\,\theta_{\max})$ with an assumed $\theta_{\max}=1^\circ$, where $\bar{\mathbf e}_j$ follows Eq.~\eqref{eq:tube} on $(A_a,B_a)$ with a disturbance torque $\bar w_\tau$ in place of $a_wT_s$. That torque is $0$ for nominal MPC, $0.01$~N\,m for the rigid tube, the gravity-gradient and model-error allowance a rigid-body designer would carry, and $\tau_{\mathrm{com}}+\tau_{\mathrm{reac}}=1.15$~N\,m from Eq.~\eqref{eq:tslosh} for the slosh-aware one. Problems \eqref{eq:ocp} and \eqref{eq:attocp} with these slosh bounds are the slosh-aware tube MPC; with the rigid allowances they are the rigid tube, and with none, nominal MPC. The attitude campaign flies $\tau_{\max}=4$~N\,m.

\subsection*{Aim-off and HD-TUBE}

A design that coasts the last $3$~m blind cannot correct the drift of that coast. Without compensation the vehicle reaches the $3$~m gate centred to $0.009$~m and drifts $0.13$~m off-centre by contact. Every controller therefore applies the same radial aim-off $a_z$ of Eq.~\eqref{eq:ocp}, set after a gain sweep to $-0.6$ times the predicted coast drift, which reduces that to a $0.015$~m miss. The aim-off scales as $1/v_{\mathrm{contact}}$, so a slower controller hands over with a larger offset.

The sixth controller, HD-TUBE, carries Hartley's horizon-selection rule \cite{Hartley2012} into the comparison. The reference architecture's horizon-dependent tube \cite{Iskender2026EllipticalTube,iskender2026horizon} evaluates the recursion \eqref{eq:tube} over whatever horizon a phase flies, whereas HD-TUBE chooses the horizon from it. Since Eq.~\eqref{eq:tube} tightens more at a longer horizon, HD-TUBE takes the longest candidate $N\in\{15,25,35,45,60,75\}$ whose free-response endpoint still lies inside the tightened corridor. On the terminal campaign it selects $N^\star=15$ on $95.33\%$ of $8354$ calls, and flying $N=15$ alone reproduces its gate offset, coast violations and propellant to three decimal places. HD-TUBE therefore measures the effect of a $45$~s horizon against the others' $75$~s.

\begin{table}[tbp]
\centering
\caption{Monte Carlo campaigns on the Basilisk plant and the dispersions each draws; seeds $42+k$ are shared across plants and controllers.}
\label{tab:montecarlo}
\footnotesize
\setlength{\tabcolsep}{3pt}
\begin{tabularx}{\columnwidth}{@{}>{\raggedright\arraybackslash}X>{\raggedright\arraybackslash}p{0.34\columnwidth}lr@{}}
\toprule
Campaign (draw), span & Flown & Plants & $n$ \\
\midrule
Mission (M), 476.8~km to contact & published form, HDE, HDE-NT & both & 20 \\
Closing leg (C), 4393~m to 100~m gate & six controllers, YA, $3^\circ$ & both & 20 \\
Terminal leg (T), 98~m to contact & six controllers, CW, $3^\circ$ & both & 100 \\
Corridor sweep (T) & six controllers, $1.5^\circ$ to $10^\circ$ & coupled & $5\times30$ \\
Attitude (A), replayed final approach & five attitude controllers, 4~N\,m & both & 100 \\
\midrule
\end{tabularx}\par
\begin{tabularx}{\columnwidth}{@{}>{\raggedright\arraybackslash}p{0.25\columnwidth}>{\raggedright\arraybackslash}X>{\raggedright\arraybackslash}p{0.12\columnwidth}@{}}
Source & Distribution, drawn once per seed & Used in \\
\midrule
Along-track start & $98+\mathcal N(0,4)$~m; $4393\,(1+\mathcal N(0,0.02))$~m & T; C \\
Cross-track, radial start & $\mathcal N(0,3)$~m; $\mathcal N(0,40)$~m; $\mathcal N(0,200)$~m about the reference hand-over state & T; C; M \\
Along-track rate & $-0.05+\mathcal N(0,0.005)$~m/s; $-0.5+\mathcal N(0,0.05)$~m/s & T; C \\
Cross-track, radial rates & $\mathcal N(0,0.005)$~m/s; $\mathcal N(0,0.05)$~m/s & T; C \\
Attitude, body rate & $\mathcal N(0,0.4^\circ)$ and $\mathcal N(0,0.02^\circ/\mathrm{s})$ per axis; zero in M & T, C, A \\
Slosh state & $\mathcal N(0,10~\mathrm{mm})$ and $\mathcal N(0,4~\mathrm{mm/s})$ per particle; zero in M & T, C, A \\
Fill multiplier & $\mathcal N(1,0.20)$ clipped to $[0.75,1.45]$ & T, C, A \\
 & none: nominal fill, 1.00 & M \\
Navigation, execution, parameters & none: every law reads the true simulator state, impulses are exact, no parameter mismatch is drawn & all \\
\bottomrule
\end{tabularx}
\end{table}

\section{Results}
\label{sec:results}

Both campaigns pair trials by seed (Table~\ref{tab:montecarlo}). Rigid- and coupled-plant runs start from the same drawn state and fill, with no navigation or execution error drawn on either, so a comparison is a median of per-trial differences, and every interval below is a $4000$-resample percentile bootstrap on that median.

\subsection*{The reference guidance on the coupled plant, homing to docking}
\label{sec:homing}

\begin{figure*}[tbp]
\centering
\includegraphics[width=\textwidth]{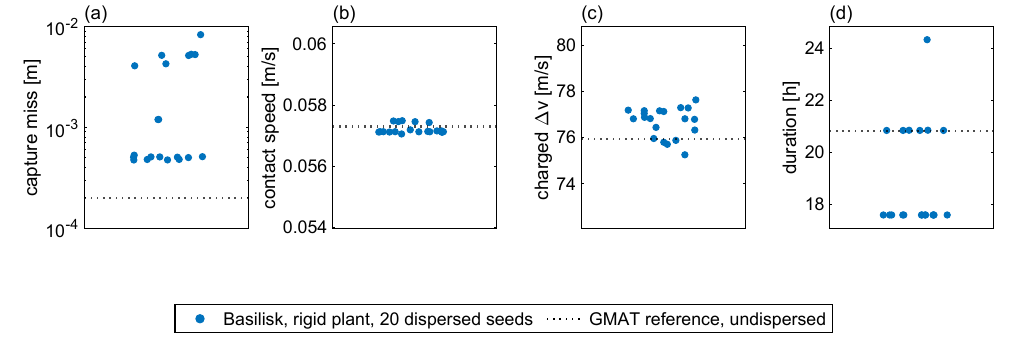}
\caption{Rigid-plant gate seed by seed, 20 dispersed Basilisk seeds (dots) against the undispersed GMAT run (dotted line): (a) miss, (b) contact speed, (c) charged $\Delta v$, (d) duration.}
\label{fig:gate}
\end{figure*}

We first check the rigid plant. Flying the published form under the $50\times50$ field on twenty dispersed seeds, Basilisk matches GMAT's undispersed run to within half a per cent in contact speed, $0.0571$ against $0.0573$~m/s, and two per cent in charged $\Delta v$, $76.82$ against $75.96$~m/s. Both miss by less than a millimetre, and the GMAT point lies inside the Basilisk spread (Fig.~\ref{fig:gate}).

\begin{figure*}[tbp]
\centering
\includegraphics[width=\textwidth]{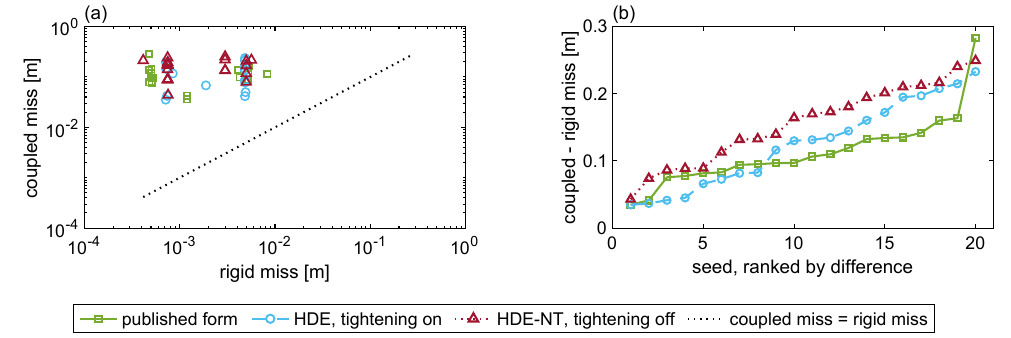}
\caption{Mission miss per paired seed, rigid against coupled plant, all three forms of Table~\ref{tab:a3homingposition}: (a) coupled against rigid miss, log--log; (b) coupled minus rigid miss, seeds ranked.}
\label{fig:sloshmission}
\end{figure*}

Table~\ref{tab:a3homingposition} then flies the three forms on both plants. Every form misses by more on the coupled plant on every one of the twenty seeds (Fig.~\ref{fig:sloshmission}), the published form least, while the propellant barely moves. One published-form seed, at $0.284$~m, falls outside the $0.20$~m capture envelope, and the other nineteen capture.

\begin{table*}[tbp]
\centering
\caption{Hub-point miss at contact, reference-guidance forms flown homing to contact: coupled minus rigid, paired median over 20 seeds, bootstrap 95\% CI.}
\label{tab:a3homingposition}
\small
\setlength{\tabcolsep}{3pt}
\resizebox{\textwidth}{!}{%
\begin{tabular}{@{}lccccccc@{}}
\toprule
Form & Rigid [m] & Coupled [m] & $\Delta$ [m] & 95\% CI & Worse on & $v_{\mathrm{contact}}$ coup.\ [m/s] & $\Delta$ charged $\Delta v$ [m/s] \\
\midrule
Published form & 0.0005 & 0.1067 & $+0.1018$ & $[+0.0886,+0.1333]$ & 20/20 & 0.0588 & $+0.26$ \\
HDE, tightening on            & 0.0048 & 0.1356 & $+0.1307$ & $[+0.0772,+0.1664]$ & 20/20 & 0.0770 & $+0.11$ \\
HDE-NT, tightening off        & 0.0008 & 0.1700 & $+0.1671$ & $[+0.1224,+0.1977]$ & 20/20 & 0.0767 & $+0.03$ \\
\bottomrule
\end{tabular}}
\end{table*}

Most of that increase comes from where the propellant sits rather than from its motion. The mission runs command no attitude on the Basilisk hub, and the propellant puts the centre of mass $63.5$~mm off the thrust line, so every lateral burn spins the coupled hub. Flown again closed loop, the published form's first seed turns $1830$ times in $20.6$~h and is still turning at $8.8^\circ$/s at contact. With the liquid frozen at the tank centre the same seed still tumbles and misses by $0.083$~m, against $0.098$~m with the liquid free. Separating the liquid's own share would take a frozen-liquid campaign, which we have not flown; the terminal-leg campaigns below hold the hub at a fixed inertial attitude and do not tumble.

\begin{figure*}[tbp]
\centering
\includegraphics[width=\textwidth]{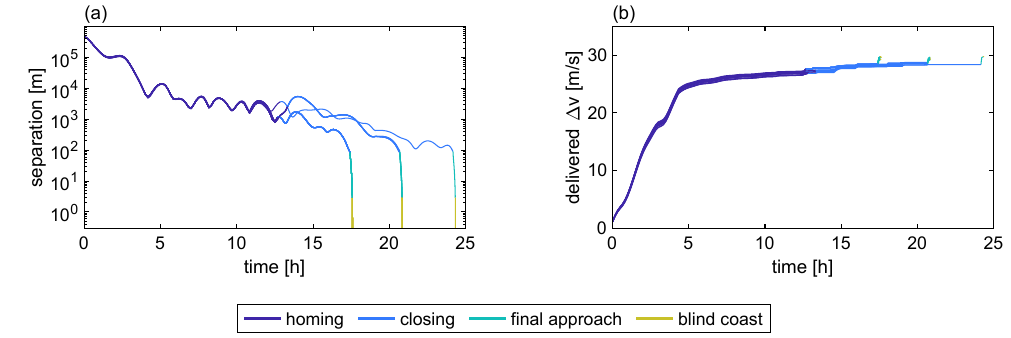}
\caption{The mission on the rigid plant, 20 dispersed seeds coloured by logged phase: (a) separation, log scale, and (b) cumulative delivered $\Delta v$, both against time.}
\label{fig:mission}
\end{figure*}

\begin{figure*}[tbp]
\centering
\includegraphics[width=\textwidth]{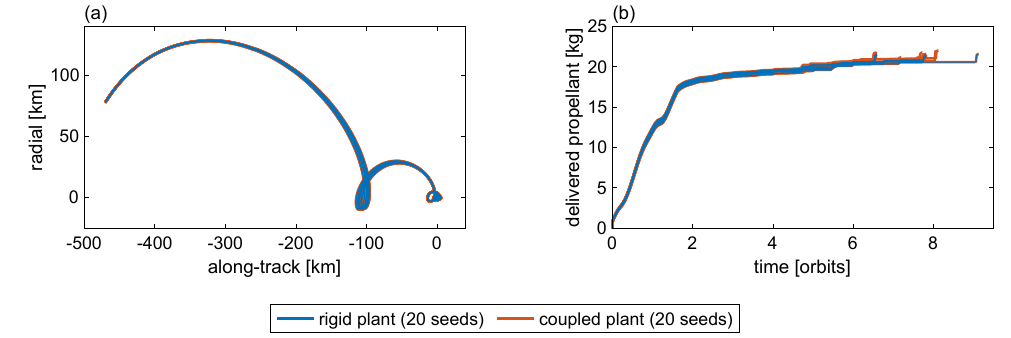}
\caption{Mission ensemble, 20 seeds per plant: (a) orbit-plane tracks, radial against along-track in km, coupled drawn over rigid; (b) propellant equivalent of the delivered impulse against time in orbits.}
\label{fig:ensemble}
\end{figure*}

\begin{figure}[tbp]
\centering
\includegraphics[width=\columnwidth]{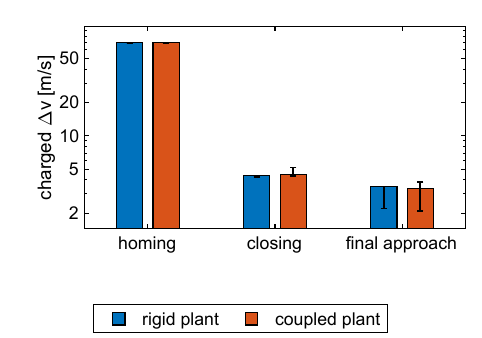}
\caption{Charged $\Delta v$ per leg, coupled against rigid: medians and interquartile bars over 20 paired seeds, log axis.}
\label{fig:stages}
\end{figure}

The tube ablation, HDE minus HDE-NT paired per seed, is inconclusive at twenty seeds. It leaves the rigid plant's miss unchanged, and on the coupled plant it lowers the median miss by $5.9$~cm on $13$ of $20$ seeds, but the interval, $[-0.102,+0.053]$~m, straddles zero. Homing carries $90\%$ of the published form's charged budget, $69.3$ of $76.8$~m/s, and the liquid moves it by $0.13$~m/s (Fig.~\ref{fig:stages}). Before the terminal leg the coupled plant shows mainly in the timing, with a median duration of $20.8$~h against $17.6$~h. More than four fifths of the delivered $\Delta v$ is spent within the first five hours (Fig.~\ref{fig:mission}), and the two plants' ensembles overlap at kilometre scale, with the propellant equivalent of the delivered impulse at a median of $21.47$~kg rigid and $21.56$~kg coupled (Fig.~\ref{fig:ensemble}).

\subsection*{Terminal leg: six controllers, 98~m to contact}

\begin{figure*}[tbp]
\centering
\includegraphics[width=\textwidth]{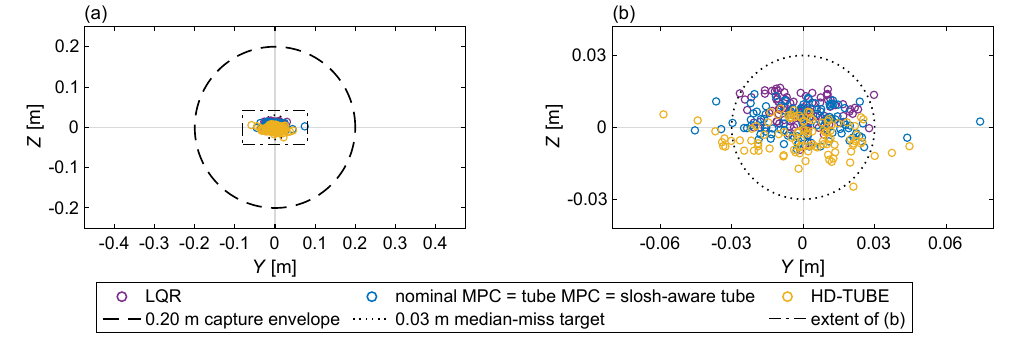}
\caption{Hub-point crossings ($Y$--$Z$) of the plane $5.65$~m along-track of the target hub, capturing controllers, 100 coupled-plant trials: (a) at the $0.20$~m capture envelope and (b) zoomed; the dotted ring is the $0.03$~m median-miss target.}
\label{fig:capture}
\end{figure*}

The six controllers fly from $98$~m along-track to contact on the coupled plant inside the $3^\circ$ corridor, $100$ paired seeds (Table~\ref{tab:position}). Miss and capture are scored at the chaser hub point against the $5.65$~m along-track standoff. PD captures on none: its $1.87$~m median miss is an order of magnitude outside the envelope, and it is kept as a floor. The other five capture on all $100$, and LQR has the tightest median miss. Nominal MPC, the rigid tube and the slosh-aware tube are identical to four significant figures and differ by at most $33$~$\mu$m on any trial, which is why Fig.~\ref{fig:capture} draws them as one series. HD-TUBE matches their miss with the lowest contact speed and the least propellant, $11.93\%$ less than nominal MPC on $95$ of $100$ seeds and $5.82\%$ less than LQR on $86$. We also swept the corridor half-angle from $1.5^\circ$ to $10^\circ$, thirty paired seeds per angle. At every angle the slosh-aware tube's median miss equals nominal MPC's to four decimals, and HD-TUBE keeps its saving.

\begin{table*}[tbp]
\centering
\caption{Terminal leg, six controllers, 98~m to contact, coupled plant, fixed inertial attitude, $3^\circ$ corridor, 100 seeds: hub-point miss medians, 95th percentile, paired change against the rigid plant, violation and capture counts.}
\label{tab:position}
\small
\setlength{\tabcolsep}{4pt}
\begin{tabular}{@{}lccccccc@{}}
\toprule
Controller & Miss med. & p95 & $\Delta$ vs.\ rigid & $v_{\mathrm{contact}}$ & Propellant & Corridor & Captured \\
 & [m] & [m] & [mm] (worse on) & [m/s] & [kg] & ctrl / coast & \\
\midrule
PD               & 1.8739 & 2.2016 & $-3.9$ (0/100)  & 1.3517 & 2.747 & 1538 / 57 & 0/100   \\
LQR              & 0.0129 & 0.0253 & $+4.8$ (94/100) & 0.0774 & 3.123 & 161 / 10  & 100/100 \\
Nominal MPC      & 0.0138 & 0.0304 & $+8.0$ (90/100) & 0.0682 & 3.351 & 156 / 114 & 100/100 \\
Tube MPC         & 0.0138 & 0.0304 & $+8.0$ (90/100) & 0.0682 & 3.351 & 156 / 114 & 100/100 \\
Slosh-aware tube & 0.0138 & 0.0304 & $+8.0$ (90/100) & 0.0682 & 3.351 & 156 / 114 & 100/100 \\
HD-TUBE          & 0.0138 & 0.0353 & $+6.6$ (94/100) & 0.0526 & 2.963 & 156 / 801 & 100/100 \\
\bottomrule
\end{tabular}
\end{table*}

The corridor counts need reading apart. Of the $100$ seeds, $32$ start outside the $3^\circ$ cone. On the $20$ trials with stored trajectories every controlled-phase violation belongs to such a seed and none follows a first entry into the cone, so the initial conditions set the controlled counts. Below the $3$~m blinding range no controller is running and the cone's half-width closes to zero, so any lateral offset at hand-over eventually crosses it. This happens most often with HD-TUBE, whose slower arrival needs a larger aim-off. The paired column gives the liquid's cost under the inertial hold, in which at least $82\%$ of each along-track burn acts across $\hat{\mathbf b}_1$ and drives the lateral slosh particles. Every capturing controller's hub-point miss grows on the coupled plant, $4.8$ to $8.0$~mm at the median and worse on $90$ to $94$ of $100$ seeds (for nominal MPC $+8.0$~mm, interval $[+6.6,+9.6]$~mm), for $2.6$~g more propellant with nominal MPC, $+0.077\%$.

The attitude campaign, a fixed-attitude hold replaying the final-approach burns without orbital motion, is where the controllers separate (Table~\ref{tab:attitude}). On the rigid plant the same five controllers settle to at most $0.0012^\circ$, so the spread on the coupled plant comes from the liquid and from the propellant's offset centre of mass moving the vehicle. There, nominal MPC and the rigid tube settle identically, and the slosh-aware tube improves on both at every statistic, on $99$ of $100$ paired seeds, for more torque impulse and four fallbacks to the ancillary law over the $100$ trials. The difference comes from the pointing constraint of Section~\ref{sec:ctrl}, which the slosh bound makes active and the rigid allowance does not. At the native $20$~N\,m the medians are unchanged, because neither authority saturates. The gain was measured on holding a non-rotating attitude, and the attitude tube was not flown on docking-axis pointing.

\begin{table}[tbp]
\centering
\caption{Settled error of a fixed-attitude hold in the attitude campaign (replay without orbital motion), coupled plant, 100 seeds, 4~N\,m torque ball: median, 95th percentile, worst and RMS, with the torque impulse spent.}
\label{tab:attitude}
\small
\setlength{\tabcolsep}{3pt}
\resizebox{\columnwidth}{!}{%
\begin{tabular}{@{}lcccccc@{}}
\toprule
Controller & Median & p95 & Worst & RMS & Impulse & Beats \\
 & [$^\circ$] & [$^\circ$] & [$^\circ$] & [$^\circ$] & [N\,m\,s] & MPC \\
\midrule
PD               & 1.8572 & 2.3534 & 2.5569 & 0.5522 & 128.5 & 0/100  \\
LQR              & 0.4269 & 0.5468 & 0.5991 & 0.1264 & 124.7 & 0/100  \\
Nominal MPC      & 0.0711 & 0.0855 & 0.0926 & 0.0202 & 127.8 & n/a    \\
Tube MPC         & 0.0711 & 0.0855 & 0.0926 & 0.0202 & 127.8 & 0/100  \\
Slosh-aware tube & 0.0620 & 0.0734 & 0.0803 & 0.0178 & 131.1 & 99/100 \\
\bottomrule
\end{tabular}}
\end{table}

\subsection*{Closing leg: 4393~m to the 100~m gate}

Over the closing leg, flown on the YA predictor, the slosh effect vanishes. PD and LQR leave the corridor, while the four MPC-family controllers reach the gate $0.0135$ to $0.0136$~m off-centre on the coupled plant against $0.0138$~m on the rigid one, with no corridor violation and $19.5$~kg of propellant either way. HD-TUBE saves nothing here. Under continuous control at $0.4$ to $2$~m/s the liquid's reaction, of order $10^{-4}$~m/s$^2$, is small against what the controller already commands, and, under the inertial attitude hold both legs fly, it shows only once the vehicle stops and coasts blind.

\subsection*{The extended design on the docking geometry}
\label{sec:litveh}

A second series flies the docking geometry with a larger tank. On the MSR chaser a sphere of $0.50$~m radius holds $330$~kg of propellant at the median hand-over ($253$ to $459$~kg across seeds), $1.0$~m from $B$, either up $\hat{\mathbf b}_3$ as in Fig.~\ref{fig:vehicle} or behind $B$ on the docking axis $\hat{\mathbf b}_1$. The liquid is carried by two particles of $160.3$~kg on a $23.4$~s mode with a linear limit of $74$~mm ($51$ to $84$~mm across loads), and it pulls the centre of mass of the $1805$~kg vehicle $18.3$~cm toward the tank. On this mode a box tube on Eq.~\eqref{eq:aug} would grow by $1.41$ per step although $\rho(A_s^d)=0.997$, so here too both arms predict with the rigid $(A,B)$ and the cost has no slosh term.

\begin{figure*}[tbp]
\centering
\includegraphics[width=\textwidth]{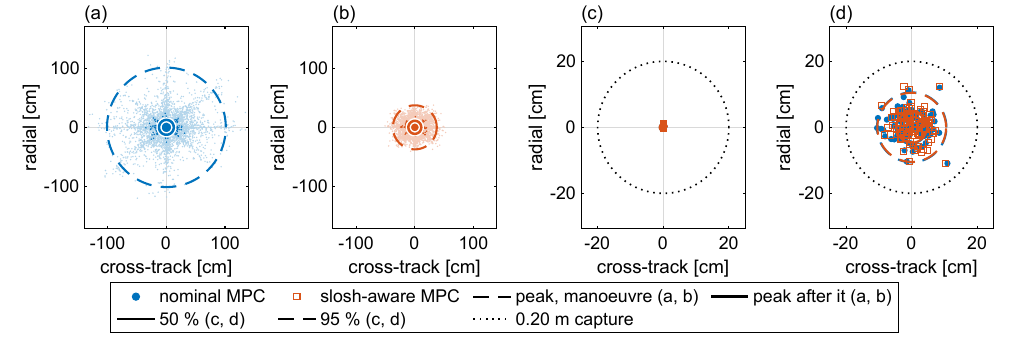}
\caption{Tank on the docking axis, $100$ paired seeds: (a, b) the beam of $\hat{\mathbf b}_1$ on a screen $100$~m ahead, light during the hand-over manoeuvre; (c, d) docking-plane crossings, circles holding $50\%$ and $95\%$; (a--c) true state, dry centre of mass unknown to $1$~cm, (d) noisy navigation.}
\label{fig:dockaxis}
\end{figure*}

The attitude reference is the target's orbital frame, with $\hat{\mathbf b}_1$ on the approach axis and $\hat{\mathbf b}_3$ radially out, and the miss is scored where a port $2.5$~m ahead of $B$ on $\hat{\mathbf b}_1$ crosses the docking plane (Fig.~\ref{fig:dockaxis}). Besides the true state, campaigns fly noisy navigation, $3$~s old, predicted to now. The relative state is measured with $5$~mm and $0.5$~mm/s errors ($1\sigma$ per axis) and reaches the controllers one $3$~s step late, and both arms predict it to now on the CW model through the burns fired since. These campaigns also disperse the plant's slosh parameters, tank position and inertia. Both arms solve Eq.~\eqref{eq:ocp} with $Q=\mathrm{diag}(0.003,100,100,800,800,800)$, $k_v=0.0167~\mathrm{s}^{-1}$ and the input rows replaced by $\mathbf u_j\in\mathcal U$, a $26$-direction polytope inscribed in $\|\mathbf u\|\le5$~mm/s. Both also fly the attitude program \eqref{eq:attocp} untightened, $\bar w_\tau=0$, with $\tau_{\max}=20$~N\,m unless Table~\ref{tab:lit} names wheels or pulsed thrusters. Slosh-aware MPC keeps the tube of Eq.~\eqref{eq:tube}, with $|B|\,\mathbf 1_3$ in place of $|B\mathbf 1_3|$, and $a_w$ of Eq.~\eqref{eq:aw} on the flown tank's particles, with $m_{\mathrm{tot}}$ the vehicle's own mass on Sloshsat and AVUM and, on the MSR chaser, the $1575$~kg of the campaigns above, $10$ to $23\%$ above the bound at the flown mass. It adds a liquid set $\mathbf u_0\in\mathcal L_k$ on the first burn, the burns after which each lateral particle's phase-plane amplitude stays below $\bar x$. For $\hat{\mathbf b}_i$, $i=2,3$, written $\mathbf c_i$ in the TOF frame,
\begin{equation}
\dot{\hat\rho}_i-\omega_{\mathrm{lo}}\sqrt{\bar x^2-\hat\rho_i^2}\le\kappa\,\mathbf c_i^\top\mathbf u_0\le\dot{\hat\rho}_i+\omega_{\mathrm{lo}}\sqrt{\bar x^2-\hat\rho_i^2},
\label{eq:liquidset}
\end{equation}
with $\kappa=m_{\mathrm{tot}}/(m_{\mathrm{tot}}-m_s)$, since a short burn changes the particle's velocity relative to the hub by $-\kappa\,\mathbf c_i^\top\mathbf u_0$. Here $\hat\rho_i$ and $\dot{\hat\rho}_i$ are an extended Kalman filter's estimates from a hub accelerometer, $\omega_{\mathrm{lo}}=f_{\mathrm{lo}}\omega_r$ with $\omega_r$ the particle frequency relative to the hub and $[f_{\mathrm{lo}},f_{\mathrm{hi}}]\subseteq[0.8,1.2]$ the band the filter has not ruled out, and $\bar x=(0.9\,x_{\mathrm{lin}}-b_k)/f_{\mathrm{hi}}^2$ is $0.9$ of the linear limit less the filter's $4.3\sigma$ amplitude bound $b_k$, at the band's upper frequency, or the current amplitude if that is larger. Zero lies in every interval, so $\mathcal L_k$ is never empty, and the component along $\hat{\mathbf b}_1$ keeps $\mathcal U$. A burn lasts a finite time, so the burn the program returns is checked against the particles' exact response and scaled down until no particle ends above the largest of $\bar x$, its amplitude before the burn and where it would end without the burn. The burn is then fired through the estimated centre of mass, $\hat{\mathbf r}_{C/B}=\bar{\mathbf r}_{C/B}+(m_s/m_{\mathrm{tot}})\hat{\bm\rho}$, the propellant's static offset plus the particles' share, with the couple $\boldsymbol\tau_{\mathrm{ff}}=\hat{\mathbf r}_{C/B}\times\mathbf F$ added. One $3$~s step of slosh-aware MPC thus estimates the liquid, bounds the first burn by Eq.~\eqref{eq:liquidset}, tightens the corridor by Eq.~\eqref{eq:tube}, solves Eq.~\eqref{eq:ocp}, checks the burn and fires it with $\boldsymbol\tau_{\mathrm{ff}}$, where nominal MPC fires through $B$.

The literature vehicles are Sloshsat FLEVO \cite{Veldman2007} and Vega's AVUM stage \cite{Reyhanoglu2012}. Sloshsat has $129$~kg with $33.5$~kg of water \cite[p.~3]{Vreeburg2005NLR} on a soft $20.8$~s mode, which we derived from the flight tank \cite[p.~7]{Vreeburg2005NLR} with Dodge's cylindrical-tank parameters \cite[Table~3.2]{Dodge2000} and which is indicative at a Bond number of $8.8$. AVUM has $1425$~kg with $450$~kg of propellant and takes its mode from \cite[Table~6.1]{RubioHervas2013}, with a $2.2$~s period shorter than the $3$~s step. One $5$~mm/s burn takes Sloshsat's water to $51\%$ of its linear limit and AVUM's to $3.4\%$. On AVUM's stiff mode the tube sized by Eq.~\eqref{eq:aw} closes the corridor and parks the vehicle a median $67$~m out, with $1$ of $100$ reaching contact, so AVUM flies a prototype tube sized on the hub's motion, at most $\mu_s x_{\mathrm{lin}}$ from the centre of mass.

\begin{table}[tbp]
\centering
\caption{Docking-geometry campaigns, nominal / slosh-aware MPC, 100 paired seeds; docking and $\Delta v$ exclude the two with navigation not predicted to now.}
\label{tab:lit}
\footnotesize
\setlength{\tabcolsep}{3pt}
\resizebox{\columnwidth}{!}{%
\begin{tabular}{@{}lccccc@{}}
\toprule
Group & Campaigns & Liquid & Peak att. & Outside & $\Delta v$ \\
 & & $>x_{\mathrm{lin}}$ & change [\%] & 0.20~m & [\%] \\
\midrule
400 kg tank above the hub & 6 & 221 / 0 & $-27$ to $-74$ & 12 / 11 of 500 & $-0.2$ to $+2.3$ \\
400 kg tank on the docking axis & 7 & 172 / 0 & $-31$ to $-64$ & 10 / 16 of 600 & $-0.2$ to $+2.0$ \\
400 kg, pulsed thrusters & 3 & 40 / 0 & $-10$ to $-24$ & 0 / 0 of 300 & $+0.4$ to $+2.5$ \\
400 kg, wheels 0.5~N\,m & 3 & 38 / 0 & $-94$ to $-99$ & 11 / 0 of 300 & $+1.7$ to $+3.3$ \\
400 kg, wheels 1~N\,m & 1 & 10 / 0 & $-82$ & 0 / 0 of 100 & $+1.6$ \\
Sloshsat FLEVO & 2 & 193 / 8 & $-22$ to $-55$ & 0 / 0 of 200 & $+5.9$ to $+14.5$ \\
AVUM, prototype tube & 2 & 0 / 0 & $-89$ to $-95$ & 0 / 0 of 200 & $-2.5$ to $-0.4$ \\
\bottomrule
\end{tabular}}
\end{table}

\begin{figure*}[tbp]
\centering
\includegraphics[width=\textwidth]{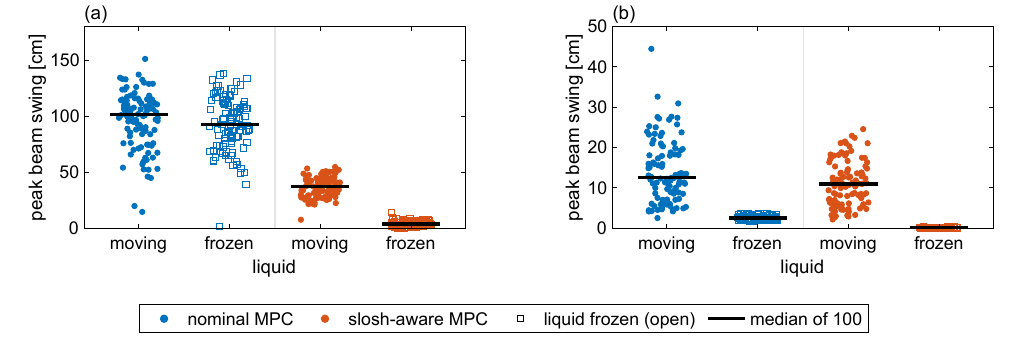}
\caption{Peak beam swing at $100$~m per approach, tank on the docking axis, true state, liquid moving and frozen on the same $100$ seeds: (a) during the hand-over manoeuvre, (b) after it.}
\label{fig:frozen}
\end{figure*}

Nominal MPC takes the liquid past its linear limit on $674$ approaches and slosh-aware MPC on $8$, all on Sloshsat and $7$ of them already past it at hand-over. On the true state that is $24$ against $0$ of $100$ with the MSR tank, and $96$ against $0$ driven over on Sloshsat. Pointing, scored as the peak attitude error from $2$~min on, improves in every campaign, by $10$ to $99\%$, and on the true state by $52$, $55$ and $95\%$ on the MSR chaser, Sloshsat and AVUM. Frozen-liquid twins give $-99.6$, $-99.2$ and $-100\%$, so the gain comes from the thrust line through the propellant-shifted centre of mass rather than from rejecting the slosh. Where the tank sits sets the size of that gain. With the tank above $B$, a burn through $B$ pitches $\hat{\mathbf b}_1$ only by its closing component, and over four navigation conditions nominal MPC's beam on a screen $100$~m ahead peaks at $3.3$ to $25$~cm while slosh-aware MPC's stays within $2.3$~cm. On the docking axis $93$ to $97\%$ of the burn impulse acts across the arm, as does the liquid's sway. From $2$~min on, slosh-aware MPC lowers the beam's RMS there by $20$ to $45\%$, to median peaks of $27$ to $54$~cm, and the attitude error by $51\%$, or $54\%$ with noisy navigation. It also keeps the liquid in where nominal MPC lets it out on $19$ of $100$ approaches. That two-minute window starts inside nominal MPC's hand-over manoeuvre, the sideways burns through $B$ that last until a median $123$~s (Fig.~\ref{fig:dockaxis}). Split at its end, nominal MPC's beam peaks at $101$~cm against slosh-aware MPC's $37$~cm during the manoeuvre and at $12.6$ against $11.0$~cm after it. With the liquid frozen (Fig.~\ref{fig:frozen}) nominal MPC still swings $93$~cm during the manoeuvre, the propellant's mass, but $2.5$~cm after it, less than with the liquid moving on $99$ of $100$ seeds, and slosh-aware MPC $3.5$ and $0.1$~cm, so the rest of its swing is the sloshing. With noisy navigation nominal MPC's sideways burns last until a median $1136$~s, and its swing remains the propellant's mass. Fired through the static centre of mass, nominal MPC swings $43$~cm in the manoeuvre, $4.0$~cm more than slosh-aware MPC on $68$ of $100$ seeds, and still lets the liquid out on $9$ of $100$ approaches. Neither attitude program models the sloshing, which has settled by contact. With noisy navigation the docking axis is then tilted $0.022^\circ$ against $0.019^\circ$, $1.0$ against $0.8$~mm at the port, beside median misses of $47.6$ and $46.5$~mm.

\begin{figure*}[tbp]
\centering
\includegraphics[width=\textwidth]{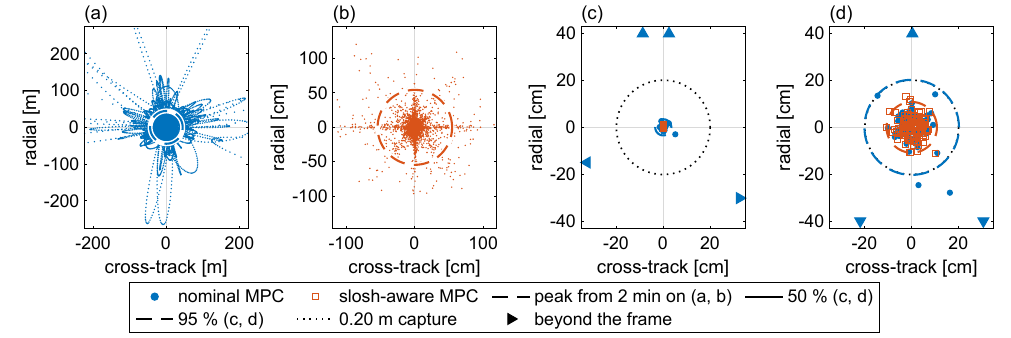}
\caption{As Fig.~\ref{fig:dockaxis} with $0.5$~N\,m reaction wheels in both arms, dry centre of mass exact, beam from $2$~min on and (b) on its own scale; triangles mark crossings beyond the frame. Only the radii of (c) differ significantly.}
\label{fig:wheels}
\end{figure*}

Docking is therefore unchanged with thruster attitude control. Over the $18$ such campaigns compared, the paired miss differs by $-1.5$ to $+3.7$~mm, significantly only with a $0.5$~m blind point ($+1.6$~mm for slosh-aware MPC). With pulsed attitude thrusters ($2.0$~N\,m\,s) the pulse limit cycle sets the tilt at contact for both, $0.18^\circ$ and $0.20^\circ$, and the miss, $15.7$ against $18.8$~mm. Docking separates once $0.5$~N\,m reaction wheels hold the attitude (Fig.~\ref{fig:wheels}). A sideways burn through $B$ carries $6.6$~N\,m about the centre of mass, $13$ times what the wheels give, so with the tank on the docking axis nominal MPC's attitude error peaks at a median $24.0^\circ$ against $0.31^\circ$, and $4$ of its $100$ approaches miss the $0.20$~m capture. Over the three wheel campaigns nominal MPC misses on $11$ of $300$ approaches and slosh-aware MPC on none, and the frozen-liquid twin of the docking-axis campaign still misses on $5$ of $100$. Slosh-aware MPC changes $\Delta v$ by $-0.4$ to $+14.5\%$ on the true state and $-2.5$ to $+5.9\%$ with noisy navigation. On the MSR chaser the most is $3.3\%$, $22$~mm/s or $0.03\%$ of the full rendezvous's $77.1$~m/s.

The plant's liquid stays linear past its limit, so the liquid bound keeps the controller's model valid without a measured performance gain. We also started the filter confident at $+0.67\,x_{\mathrm{lin}}$ with the particles at $-0.67\,x_{\mathrm{lin}}$ (tank above $B$, true state). Slosh-aware MPC then lets the liquid past its limit on $42$ of $100$ approaches, all within $24$~s, against nominal MPC's $62$. Its estimate is within half the limit after a median $48$~s and within a tenth of it after $249$~s, and pointing ($0.13^\circ$ against $0.27^\circ$) and docking hold for $8.5\%$ more $\Delta v$. With the estimate mirrored throughout, as a sign-convention error would leave it, the liquid goes past the limit on all $100$ approaches, with pointing at $0.76^\circ$ and a $65.5$~mm median miss against $3.6$~mm.

\section{Discussion}
\label{sec:discussion}

Over the last $98$~m, with the hub held at a fixed inertial attitude, every capturing controller loses $4.8$ to $8.0$~mm of median hub-point miss to the liquid for under three grams of propellant, and tube or none, all five lose a comparable amount, so none rejects the disturbance in translation. Partial tightening changes the flown trajectory only where the tightened constraint becomes active, and in attitude it does. The slosh-sized tube pulls the predicted pointing budget below the error nominal MPC settles to, so the program must then act on the slosh torque, and the settled error of the fixed-attitude hold falls by $12.8\%$.

In translation the same bound widens the lateral tube to $20.5$~mm at $N=25$, against $0.13$~mm for the rigid allowance. The corridor's half-width, however, is $5.1$~m at the $98$~m hand-over and still $0.16$~m at the $3$~m blinding point, while the controlled trajectory is centred to millimetres. The tightened corridor therefore stays slack and adds a margin in the prediction without certifying the flown cone. On the closing leg the coupled-minus-rigid difference is resolved but amounts to $0.2$~mm on a $13.5$~mm gate offset, so, under the inertial hold both legs fly, a slosh-sized margin matters only once the vehicle coasts blind and nearly stopped.

\section{Conclusions}
\label{sec:conclusions}

We made a tube MPC slosh-aware without adding slosh states to its predictor. Its tightening is sized from the reaction force and torque the propellant can exert within its analogue's linear limit, and applied only to the corridor and a pointing budget about a fixed inertial attitude. On the coupled Basilisk plant the reference rendezvous guidance loses accuracy, from $0.0005$ to $0.1067$~m of median miss over twenty paired seeds for $0.26$~m/s more charged $\Delta v$, mostly because its uncontrolled hub tumbles about the propellant's offset centre of mass. In a replay without orbital motion, the slosh-aware tube lowers the settled error of a fixed-attitude hold by $12.8\%$ at the median and $14.1\%$ at the ninety-fifth percentile against nominal and rigid-tube MPC, on $99$ of $100$ seeds, and the liquid's effect on the closing leg, flown under the same inertial hold, is negligible.

On the docking geometry, over $24$ campaigns on the MSR chaser with a $400$~kg tank, Sloshsat FLEVO and AVUM, the extended slosh-aware MPC lets the liquid past its linear limit on $8$ of the $2400$ approaches, against $674$ with nominal MPC. Pointing improves by $52$, $55$ and $95\%$ on the three vehicles and in every campaign, and the gain persists with the liquid frozen, so it comes from knowing where the propellant sits. Docking is as accurate with thruster attitude control and more reliable with $0.5$~N\,m reaction wheels, $300$ of $300$ approaches against $289$, for at most $3.3\%$ more $\Delta v$ on the MSR chaser. With an ideal thruster model the design also transfers to both literature vehicles, AVUM with a prototype momentum tube.

Neither recursive feasibility nor asymptotic stability is claimed for any controller. Two questions remain open. A frozen-liquid baseline would separate the liquid's own share of the mission miss, and a tube ablation on the coupled plant needs about a hundred paired seeds.

\subsection*{Limitations and future work}

The plant is an equivalent mechanical model, validated like most slosh models almost entirely in 1\,g, and it holds only below its linear limit, $43$~mm on the $100$~kg tank (above a fill of $0.70$) and $74$~mm at the $400$~kg tank's median load. In the docking-geometry design, $3$~s old navigation not predicted to now leaves the lateral loop marginally unstable (spectral radius $1.0031$), so docking is compared only with it predicted to now. With pulsed attitude thrusters docking is a tie, so the docking gain needs weak attitude authority, and the plant fires an ideal force in any direction, more lateral freedom than Sloshsat's and AVUM's thrusters give. Natural extensions are real-time identification of the nonlinear slosh parameters, hardware-in-the-loop tests that take a planar free surface from 1\,g towards 0\,g, and coupling with flexible appendages \cite{IskenderJGCD2026}.

\section*{Acknowledgements}
The authors thank the developers of OSQP \cite{Stellato2020}, Basilisk \cite{Kenneally2020} and GMAT \cite{Hughes2014GMAT}, whose tools underpin every result.

\bibliography{sloshing_aware_tube_mpc}

\end{document}